\documentclass{article}
\usepackage{geometry} 
\usepackage{amsmath}
\usepackage{amssymb}
\usepackage{accents}
\usepackage{latexsym}
\usepackage{authblk}
\usepackage{hyperref}
\usepackage{bmpsize}
\usepackage[square,numbers]{natbib}  
\usepackage{appendix}
\usepackage{color} 
\hypersetup{
setpagesize=false,
 bookmarksnumbered=true,%
 bookmarksopen=true,%
 colorlinks=true,%
 linkcolor=blue,
 citecolor=red,
}
\makeatletter
 
  \@addtoreset{equation}{section}
\makeatother

\title{Cartan $F(R)$ Gravity and Equivalent Scalar-Tensor Theory}

\author[1,2,3,4]{Tomohiro Inagaki }
\author[4]{Masahiko Taniguchi}

\affil[1]{Information Media Center, Hiroshima University, Higashi-Hiroshima, 739-8521,Japan}
\affil[2]{Core of Research for the Energetic Universe, Hiroshima University,Higashi-Hiroshima, 739-8526, Japan}
\affil[3]{Lab. Theor. Cosmology, Tomsk State University of Control Systems and Radioelectronics (TUSUR), 634050 Tomsk, Russia}
\affil[4]{Graduate School of Advanced Science and Engineering, Hiroshima University, Higashi-Hiroshima, 739-8526, Japan }

\date {} 

\begin{document}
\maketitle

\begin{abstract}
We investigate the Cartan formalism in $F(R)$ gravity. 
$F(R)$ gravity has been introduced as a theory to explain cosmological accelerated expansion by replacing the Ricci scalar $R$ in the Einstein-Hilbert action with a function of $R$.
As is well-known, $F(R)$ gravity is rewritten as a scalar-tensor theory by using the conformal transformation.
Cartan $F(R)$ gravity is described based on the Riemann-Cartan geometry formulated by the vierbein.
In the Cartan formalism, the Ricci scalar $R$ is divided into two parts, one derived from the Levi-Civita connection and the other from the torsion.
Assuming the spin connection independent matter action, we have successfully rewritten the action of Cartan $F(R)$ gravity into the Einstein-Hilbert action and a scalar field with canonical kinetic and potential terms without any conformal transformations.
The resulting scalar-tensor theory is useful in applying the usual slow-roll scenario.
As a simple case, we employ the Starobinsky model and evaluate fluctuations in the cosmological microwave background radiation.
\end{abstract}

\section{Introduction}
Cartan formalism is a natural extension of the General Relativity (GR) based on the Riemann-Cartan geometry formulated by the basis called vierbein~\cite{cartan:1923}. 
Einstein introduced it to unify gravity and electromagnetism in 1928~\cite{Goenner:2004se}. 
Later, the Cartan formalism became known as the Einstein-Cartan-Kibble-Sciama (ECKS) theory in the 1960s~\cite{SCIAMA:1964rmp,Kibble1961jmp}. 
ECSK theory is currently being studied for applications to cosmological problems~\cite{Boehmer:2008ah,Poplawski:2010kb,Magueijo:2012ug,Shaposhnikov:2020frq,Shaposhnikov:2020gts,Iosifidis:2021iuw,Cabral:2021dfe,Piani:2022gon}.

Another extension of GR is to introduce a term that modifies the Einstein-Hilbert action.
$F(R)$ gravity is the theory in which the Ricci scalar, $R$, in the Einstein-Hilbert action is replaced by an arbitrary function of Ricci scalar, $F(R)$~\cite{Buchdahl:1983zz,Nojiri:2017ncd}.
The $F(R)$ gravity can be rewritten as an equivalent scalar-tensor theory through the conformal transformation~\cite{Nojiri:2017ncd,jordan1955schwerkraft}.
It should be noticed that there is some discussion about the equivalence of physics before and after the conformal transformation~\cite{Catena:2006bd,Steinwachs:2011zs,Kamenshchik:2014waa,Hamada:2016onh}.

One of the most famous $F(R)$ gravity models is the Starobinsky model~\cite{Starobinsky:1979ty}.
After conformal transformation, the model is described by the scalar-tensor theory, whose scalar field has a potential with a flat plateau at high energy.
The potential energy accelerates the expansion of the universe while it dominates the energy density of the universe, 
As is well-known, the accelerating expansion in the early universe can solve the horizon and flatness problems~\cite{Sato:1980yn,Guth:1980zm}. 

The main purpose of this study is to adapt the Cartan formalism to the $F(R)$ gravity and apply the result to the Starobinsky-like model.
A characteristic property of Cartan $F(R)$ gravity is that the torsion does not vanish~\cite{Montesinos:2020pxv}.
The Ricci scalar $R$ is then divided into the usual part obtained from the Levi-Civita connection and the additional part obtained from the torsion.

The non-vanishing torsion is also a feature of the Palatini approach to the modified gravity and more general metric-affine geometry.
Several studies have been done on the basic properties of the metric-affine $F(R)$ gravity and its application to cosmology  ~\cite{Sotiriou:2006mu,Sotiriou:2006qn,Iosifidis:2018zjj}. 
It is found that the metric-affine $F(R)$ and Palatini $F(R)$ gravity are rewritten as a certain class of Brans-Dicke type scalar-tensor theories after a conformal transformation~\cite{Capozziello:2007tj,Capozziello:2008yx,Sotiriou:2009xt,Capozziello:2009mq,Olmo:2011uz}.

This paper shows that a class of Cartan $F(R)$ gravity obtains an Einstein-Hilbert action and a scalar field action with canonical kinetic and potential by simple scalar field's definition.
The Cartan $F(R)$ gravity does not require conformal transformation, which is different from conventional F(R) gravity theories.
Therefore, we can evaluate the Cartan $F(R)$ gravity by using the equivalent scalar-tensor theory in the original frame.
Finally, Starobinsky's potential can be derived from the Cartan $F(R)$ gravity model.
The Cartan $F(R)$ gravity is useful to consider the usual slow-roll scenario.

\section{Cartan $F(R)$ gravity}
The $F(R)$ gravity is reformulated on the Riemann-Cartan geometry described by the vierbein and the spin connection. We call it Cartan $F(R)$ gravity. The fundamental elements of the Cartan $F(R)$ gravity are the vierbein and the spin connection. Since it is a natural extension of the $F(R)$ gravity, we expect that the additional contribution to GR is described as a scalar field theory, similar to the F(R) gravity.

Here we briefly introduce the vierbein and the spin connection.
The vierbein, ${e^i}_\mu$, is defined to act as a mediator from a flat spacetime metric, $\eta_{ij}$ to a curved one, $g_{\mu\nu}$. It contains flat and curved spacetime indices and satisfies
\begin{align}\label{Eq:Definition of tetrad}
g_{\mu\nu}=\eta_{ij}{e^i}_\mu{e^j}_\nu.
\end{align}
The Affine connection is expressed as 
\begin{align*}
{\Gamma^\rho}_{\mu\nu}= {e_a}^\rho D_\nu {e^a}_\mu,
\end{align*}
with
\begin{align*}
D_\nu {e^k}_\mu=\partial_\nu {e^a}_\mu+{{\omega^k}_{l\nu}}{e^l}_\mu.
\end{align*}
where $D_\nu$ is the covariant derivative for the local Lorentz transformation and ${\omega^{ij}}_\nu$ denotes the spin connection.
The Affine connection is not necessary to be invariant under the replacement of the lower indices, ${\Gamma^\rho}_{\mu\nu}\neq{\Gamma^\rho}_{\nu\mu}$.
Thus, the geometric tensor, ${T^{\rho}}_{\mu\nu}$, called torsion arises, 
\begin{align*}
{T^{\rho}}_{\mu\nu}\equiv{\Gamma^\rho}_{\mu\nu}-{\Gamma^\rho}_{\nu\mu}.
\end{align*}
The Riemann tensor is expressed by the spin connection,
\begin{align*}
{R^{ij}}_{\mu\nu}=\partial_{\mu}{\omega^{ij}}_\nu-\partial_{\nu}{\omega^{ij}}_\mu+{\omega^i}_{k\mu}{\omega^{kj}}_\nu-{\omega^i}_{k\nu}{\omega^{kj}}_\mu,
\end{align*}
and the Ricci scalar by the spin connection and the vierbein
\begin{align*}
R={e_i}^\mu{e_j}^\nu {R^{ij}}_{\mu\nu}(\omega,\partial\omega).
\end{align*}

The action of the Cartan $F(R)$ gravity is defined by replacing the Ricci scalar $R$ in the Einstein–Cartan theory with a function of $R$,

\begin{align}\label{eqs:action FR}
S=\int d^4xe\left( \frac{{M_{\rm Pl}}^2}{2}F(R)+\mathcal{L}_{\rm{m}}\right).
\end{align}
where $M_{\rm Pl}$ indicates the Planck scale and the volume element is given by the determinant of the vierbein, $e$.
$\mathcal{L}_{\rm{m}}$ shows the Lagrangian density for the matter field. As a simple case, we assume that the matter filed does not depend on the spin connection.
The equation of motion can be derived by the variation of the action with respect to the vierbein.
\begin{align}\label{eqs:EoM FR}
F'{R^i}_\mu-\frac{1}{2}{e^i}_\mu F(R)={M_{\rm Pl}}^{-2}{\Sigma^i}_\mu,
\end{align}
where $F'(R)$ denotes the derivative of $F(R)$ with respect to $R$ and ${\Sigma^i}_\mu$ is the energy-momentum tensor of the matter filed.
From the trace of Eq.(\ref{eqs:EoM FR}), the Ricci scalar can be represented as a function, $R(\Sigma)$, where $\Sigma$ is the trace of the energy-momentum tensor.
In the absence of the matter, the Rich scalar is uniquely determined and constant except for $F(R)=\alpha R^2$~\cite{Iosifidis:2018zjj,Capozziello:2007tj,Capozziello:2008yx,Sotiriou:2009xt,Capozziello:2009mq,Olmo:2011uz}.
Hereafter, $R(\Sigma)$ is abbreviated as $R$.

The Cartan equation is derived by the variation of the action with respect to the spin connection.
The details of the derivation of the Cartan equation are given in the appendix.
The Cartan eqution is
\begin{align}\label{eqs:cartan eq in FR}
({T^\mu}_{kl}-{e_l}^\mu T_k+{e_k}^\mu T_l)F'(R)+({e_k}^\alpha{e_l}^\mu-{e_k}^\mu{e_l}^\alpha)\partial_{\alpha}F'(R)=0.
\end{align}
Since the matter filed is spin connection independent, the Cartan equation does not include a matter term.
From the Cartan equation, the torsion is represented by the derivative of $F(R)$ and the vierbein,
\begin{align}\label{eqs:torsion of FR}
{T^k}_{ij}=\frac{1}{2}({\delta^k}_j{e_i}^\lambda-{\delta^k}_i{e_j}^\lambda)\partial_\lambda\ln F'(R).
\end{align}
It should be noted that the torsion vanishes in the Einstein–Cartan theory, $F(R)=R$.
The contribution from the torsion is induced in the Cartan $F(R)$ modified gravity.
It is extracted from the Ricci scalar,
\begin{align}\label{eqs:Ricci scalar into non and torsion2}
R=R_E+T-2{\nabla_E}_\mu T^\mu,
\end{align}
where the subscript $E$ in $R_E$ and $\nabla_E$ stands for the Ricci scalar and the covariant derivative given by the Levi-Civita connection,
\begin{align*}
{{(\Gamma_E)}^\lambda}_{\mu\nu}= \frac{1}{2}g^{\lambda\rho}(\partial_\mu g_{\nu\rho}+\partial_\nu g_{\rho\mu}-\partial_\rho g_{\mu\nu}).
\end{align*}
$T_\mu$ represents the torsion vector, $T_\mu={T^\lambda}_{\mu\lambda}$, and
$T$ the torsion scalar defined by
\begin{align*}
T=\frac{1}{4}T^{\rho\mu\nu}T_{\rho\mu\nu}-\frac{1}{4}T^{\rho\mu\nu}T_{\mu\nu\rho}-\frac{1}{4}T^{\rho\mu\nu}T_{\nu\rho\mu}-T^\mu T_\mu.
\end{align*}
Thus, the Ricci scalar is divided into two parts, $R_E$ and the additional part derived from the torsion.
Substituting Eq.(\ref{eqs:torsion of FR}) into Eq.(\ref{eqs:Ricci scalar into non and torsion2}),
the additional part is represented as
\begin{align}\label{eqs:Ricci scalar to einstein and torsion}
R=R_E-\frac{3}{2}\partial_{\lambda}\ln F'(R)\partial^{\lambda}\ln F'(R)-3\Box\ln F'(R).
\end{align}

Below we consider the class of the Cartan $F(R)$ gravity expressed as $F(R)=R+f(R)$. 
From Eq.(\ref{eqs:Ricci scalar to einstein and torsion}) the gravity part of the action is rewritten as
\begin{align} \nonumber
S&=\int d^4x e\frac{{M_{\rm Pl}}^2}{2}\left(R+f(R)\right)
\\ \label{eqs:fr action Disassembled}
&=\int d^4x e\frac{{M_{\rm Pl}}^2}{2}\left(R_E-\frac{3}{2}\partial_{\lambda}\ln F'(R)\partial^{\lambda}\ln F'(R)-3\Box\ln F'(R)+f(R)\right).
\end{align}
The third term in the integrand is in the form of a total derivative.
It is assumed that $F'(R)$ vanishes at a distance and the term proportional to $\Box\ln F'(R)$ is dropped.
We introduce a scalar field $\phi$ by the replacement,  
\begin{align}\label{eqs:def scalar field}
\phi\equiv-\sqrt{\frac{3}{2}}{M_{\rm Pl}}\ln F'(R).
\end{align}
Then the second term in the integrand of the action (\ref{eqs:fr action Disassembled})
takes a form of the canonical kinetic term for the scalar field and the potential term is generated from the last term in the integrand,
\begin{align} \label{eqs:action for R and scalar}
S=\int d^4x e\left(\frac{{M_{\rm Pl}}^2}{2}R_E-\frac{1}{2}\partial_{\lambda}\phi \partial^{\lambda}\phi-V(\phi)\right),
\end{align}
where the potential, $V(\phi)$, is given by
\begin{align} \label{eqs:potential for phi in general}
V(\phi)=-\frac{{M_{\rm Pl}}^2}{2}\left. f(R)\right|_{R=R(\phi)}.
\end{align}
The potential is expressed as a function of the scalar field $\phi$ through $R=R(\phi)$.
The explicit expression for the potential is found in terms of the scalar field by solving Eq.(\ref{eqs:def scalar field}).
It is more convenient to consider the exponential form of the equation,
\begin{align}\label{eqs:relation fr and phi}
f'(R)=e^{-\sqrt{\frac{2}{3}}\frac{\phi}{{M_ {\rm Pl}}}}-1.
\end{align}

Therefore, the action (\ref{eqs:fr action Disassembled}) reduces to the scalar-tensor theory (\ref{eqs:action for R and scalar}) without any conformal transformation.
It should be noticed that the potential $V(\phi)$ is different from the one in the scalar-tensor theory obtained from the usual $F(R)$ gravity after the conformal transformation~\cite{Nojiri:2017ncd}. 

\section{$R^2$ model}
We would like to apply the derivation of the scalar-tensor theory to a specific model.
We employ a Cartan $F(R)$ gravity with a $R^2$ correction term as a simple prototype model.
The $R^2$ term gives a small correction in weakly curved spacetime. 
We introduce a mass scale $M$ and set $F(R)=R-R^2/M^2$, $f(R)=-R^2/M^2$.
It is noted that the sign of the modification term, $R^2$ is negative, opposite to the Starobinsky model. 

For $f(R)=-R^2/M^2$ the action of the equivalent scalar-tensor theory (\ref{eqs:action for R and scalar}) is given by
\begin{align} \nonumber
S&=\int d^4x e\left(\frac{{M_{\rm Pl}}^2}{2}R_E-\frac{1}{2}\partial_{\lambda}\phi \partial^{\lambda}\phi-\frac{{M_{\rm Pl}}^2}{2}\frac{R^2}{M^2}\right),
\end{align}
where we assume that the induced scalar field, $\phi$, dominates the energy density of the universe and ignores the matter action. 
From Eq.(\ref{eqs:relation fr and phi}) the Ricci scalar, $R$, is described as 
\begin{align}\label{eqs:starobinsky r}
R=\frac{M^2}{2}\left(1-e^{-\sqrt{\frac{2}{3}}\frac{\phi}{{M_ {\rm Pl}}}} \right).
\end{align}
Thus, the potential, $V(\phi)$, is fixed in terms of the scalar field,
\begin{align}\label{eqs:starobinsky potential}
V(\phi)=\frac{{M_{\rm Pl}}^2M^2}{8}\left(1-e^{-\sqrt{\frac{2}{3}}\frac{\phi}{{M_ {\rm Pl}}}} \right)^2.
\end{align}

The potential (\ref{eqs:starobinsky potential}) reproduces the one induced by the conformal transformation in the usual Starobinsky model with  $F(R)=R+R^2/ M^2$~\cite{Starobinsky:1979ty}.
In other words, the potential in the Cartan $f(R)=-R^2/M^2$ gravity is identical to that of the Starobinsky model.
If we adopt this model to the slow-roll scenario of inflation, we can regard the scalar field as the inflaton field.
The slow-roll takes place in the usual Friedmann-Robertson-Walker background.
The inflation develops the fluctuations of the cosmic microwave background with the spectral index $n_s$ and the scalar-tensor ratio $r$,
\begin{align*}
&n_s\simeq 1-\frac{2}{N},
\\
&r\simeq\frac{12}{N^2},
\end{align*}
where $N$ is the e-folding number. 
These results are of course identical to the Starobinsky model and consistent with the current observations~\cite{Aghanim:2018eyx}. It means that the primordial inflation is induced in the $R^2$ model of the Cartan $F(R)$ gravity.

\section{Conclusion}
The Cartan $F(R)$ gravity, an extension of the $F(R)$ gravity on the Riemann-Cartan geometry, has been studied.
We have constructed a derivation of the scalar-tensor theory from the Cartan $F(R)$ gravity.
The canonical kinetic term of a scalar field is induced by extracting the torsion from the Ricci scalar.
The potential term is derived from the modified gravity action, $f(R)$.
Since the derivation does not require a conformal transformation, it is free from the equivalence problem between Jordan and Einstein frames for the scalar-tensor theory derived through the conformal transformation in $F(R)$ gravity.

In contrast to previous works, the derived scalar-tensor theory can be applied to usual slow-roll scenarios in the process of the inflation.

The derivation has been adapted to a simple model with a $R^2$ correction term.
The explicit expression of the scalar-tensor theory has been derived for the model without changing the frame.
It is found that the derived scalar-tensor theory has identical potential to the Starobinsky model. 
Therefore, we have succeeded in finding a Cartan $F(R)$ gravity model to induce the inflation consistent with the current observations of the cosmic microwave background fluctuations.

In this paper, we have focused on the gravity part of the theory and assumed that the matter action does not depend on the spin connection.

The Cartan $F(R)$ gravity is associated with matter fields through the local Lorentz invariance.
It is known that torsion is also affected by matter fields.
The original ECKS theory is promoted to introduce a spinor field in a gravity theory~\cite{Hehl:1974cn,Kerlick:1975tr,Gasperini:1986mv}.
In the ECKS theory with a spinor field, a non-vanishing torsion introduces a four-fermion interaction called spin-spin interaction or Dirac-Heisenberg-Ivanenko-Hehl-Datta four body fermi interaction~\cite{Hehl:1971qi,Boos:2016cey}.
ECKS theory is also studied with the vector fileds~\cite{deAndrade:1997cj,Nieh:2017ijh} and Rarita–Schwinger spin $3/2$ fields~\cite{Rarita:1941mf}.

It is interesting to apply the scalar-tensor theory obtained from Cartan $F(R)$ gravity to the dark energy problem by analogy with the quintessence. The origin of dark energy has been investigated in conventional $F(R)$ gravity theories~\cite{Capozziello:2003tk,Copeland:2006wr}.
Geometrically induced interactions between matter fields have been also studied to explain the dark energy~\cite{Poplawski:2011wj}. 
It indicates that the Cartan F(R) gravity can explain early and late-time accelerating expansion.
We hope to extend the derivation to the model with matter fields and report the results in the future.

\section*{Acknowledgements}
For valuable discussions, the authors would like to thank Y.~Matsuo, H.~Sakamoto, and Y.~Sugiyama.

\appendix
\section{Introduction of Cartan equation}
The Cartan equation in Cartan $F(R)$ gravity can be introduced by the variation of the action in terms of spin connection, $\omega$.
First, variation of the Riemann tensor in terms of spin connection can be derived as follows,
\begin{align}
\delta {R^{i j}}_{\mu \nu}=\left({\delta^{\alpha}}_{\mu} {\delta^{\beta}}_{\nu}-{\delta^{\alpha}}_{\nu} {\delta^{\beta}}_{\mu}\right) D_{\alpha} {\delta \omega^{i j}}_{\beta}.
\end{align}
where $D_\alpha$ is the covariant derivative under the local Lorentz transformation. 
Thus one of the Ricci scalar is also
\begin{align}
\delta(e R) &=e\left({e_{i}}^{\alpha} {e_{j}}^{\beta}-{e_{i}}^{\beta} {e_{j}}^{\alpha}\right) D_{\alpha} \delta {\omega^{i j}}_\beta.
\end{align}
Finally, variation of the action in Cartan $F(R)$ gravity becomes
\begin{align} \nonumber
\delta S_{F(R)}&=\int d^{4} x \delta(e F(R))
\\ \nonumber
&=\int d^{4} x F'(R) \delta(e R)
\\ \nonumber
&=\int d^{4} x F'(R) e\left({e_{i}}^{\alpha} {e_{j}}^{\beta}-{e_{i}}^{\beta} {e_{j}}^{\alpha}\right) D_{\alpha} \delta {\omega^{i j}}_\beta
\\ \nonumber
&=\int d^{4} x -\delta {\omega^{i j}}_\beta D_{\alpha}\left[ e\left({e_{i}}^{\alpha} {e_{j}}^{\beta}-{e_{i}}^{\beta} {e_{j}}^{\alpha}\right)F'(R)\right] 
\\ \label{eqs:var of FR}
&=\int d^{4} x -\delta {\omega^{i j}}_\beta e \left[({T^\beta}_{ij}-{e_j}^\beta T_i+{e_i}^\beta T_j)F'(R)+({e_i}^\alpha{e_j}^\beta-{e_i}^\beta{e_j}^\alpha)\partial_{\alpha}F'(R)\right].
\end{align}
In the last line of Eq.(\ref{eqs:var of FR})  we need the relation as follows,
\begin{align}
 D_{\alpha}\left[ e\left({e_{i}}^{\alpha} {e_{j}}^{\beta}-{e_{i}}^{\beta} {e_{j}}^{\alpha}\right)\right]=e\left({T^{\beta}}_{ij}-{e_{j}}^{\beta}T_i+{e_{i}}^{\beta}T_j\right)
\end{align}
Then $\delta S_{F(R)}=0$
We can derive the Cartan eqution~(\ref{eqs:cartan eq in FR}),
\begin{align}\label{eqs:cartan eq apd}
({T^\mu}_{kl}-{e_l}^\mu T_k+{e_k}^\mu T_l)F'(R)+({e_k}^\alpha{e_l}^\mu-{e_k}^\mu{e_l}^\alpha)\partial_{\alpha}F'(R)=0.
\end{align}
It should be noted that Eq.(\ref{eqs:cartan eq apd}) does not coincide with that in the metric-affine $F(R)$ gravity~\cite{Sotiriou:2006mu}. 
However, the equivalent equation can be obtained by imposing a special condition on the connection~\cite{Capozziello:2007tj}.
This is because the conditions are given by Riemann-Cartan geometry.

\bibliography{ref}
\end{document}